\documentclass[prd,aps,eqsecnum,showpacs]{revtex4}

\usepackage{amsmath,amssymb}
\usepackage{bm}

\def\prez{{}_{(0)}}

\def\pret{{}_{(2)}}

\def\postz{{}_{(0)}}
\def\postt{{}_{(2)}}

\begin{document}

\title{Gradient expansion approach to nonlinear superhorizon perturbations II\\
-- a single scalar field --}

\author{Yoshiharu {\scshape TANAKA}\footnote{E-mail: yotanaka${}_{-}$AT${}_{-}$yukawa.kyoto-u.ac.jp} and Misao {\scshape SASAKI}\footnote{E-mail:  misao${}_{-}$AT${}_{-}$yukawa.kyoto-u.ac.jp}
}

\affiliation{%         %Affiliation, neglected when [addenda] or [errata]
Yukawa Institute for Theoretical Physics, Kyoto University, 
Kyoto 606-8502, 
Japan
}

%\date{\today}

%%%%%%%%%%%%%%%%%%%%%
\begin{abstract}
We formulate nonlinear perturbations of a scalar field dominated universe 
on super-horizon scales. We consider the case of a single scalar field.
We take the gradient expansion approach.
We adopt the uniform Hubble slicing and derive the general
solution valid to $O(\epsilon^2)$, where $\epsilon$ is the expansion parameter
associated with a spatial derivative, which includes both the scalar and
tensor modes. In particular, the $O(\epsilon^2)$ correction terms to
the nonlinear curvature perturbation, which become important in 
models with a non-slowroll stage during inflation, are explicitly obtained. 
\end{abstract}
%%%%%%%%%%%%%%%%%%%%%

%----------------------------
\hfill YITP-07-31
%----------------------------

\maketitle

%%%%%%%%%%%%%%%%%%%%%%%%%%%
%%%  Sec. I 
%%%%%%%%%%%%%%%%%%%%%%%%%%%
\section{Introduction}

The cosmic microwave background (CMB) anisotropies recently observed by WMAP
strongly support the inflationary cosmology, and theoretical predictions of
various models of inflation seem to be well estimated by linear perturbation
theory, with the cosmological perturbations generated
from quantum vacuum fluctuations which are well approximated by
Gaussian random fields~\cite{WMAP3y}.
Nevertheless, there has been a growing interest in detection of
possible derivations from the Gaussian statistics.
It was suggested that a deviation from the Gaussian stastics may be used to
 distinguish models of inflation~\cite{Bartolo:2004if}, and 
it may indeed be detected by PLANCK~\cite{Komatsu:2001rj} in the near future.
Thus, making clear predictions on the non-Gaussianity from inflation 
have become one of the urgent issues of the inflationary cosmology.
Since the nonlinearity is essential for the generation of non-Gaussian
perturbations, it is necessary to develop a nonlinear cosmological perturbation
theory to evaluate the non-Gaussianity from inflation.

Our goal is to formulate a nonlinear theory with which we can
calculate non-Gaussianities from any models of inflation.
The traditional approach is to develop a second-order perturbation
theory~\cite{SecondOrder,Acquaviva:2002ud,Maldacena:2002vr,NGmodels1,Seery:2005wm}.
However, here we adopt a different one, the gradient expansion approach,
in which nonlinear perturbations are solved by invoking 
spatial gradient expansion under the assumption that spatial
derivatives are small compared to time derivatives. Technically,
we introduce an expansion parameter, $\epsilon$, and associate it with
each spatial derivative. Then we expand field equations in terms of
$\epsilon$ and solve them order by order in $\epsilon$ iteratively.
The gradient expansion approach has been developed and studied by many authors 
previously~\cite{Lifshitz:1963ps,Belinsky:1982pk,Muller:1989rp,
Khalatnikov:2003ac,antiNewton,Tomita:1975kj,Salopek:1990jq,Comer:1994np,
Deruelle:1994iz,Iguchi,Salopek:1990mp,HJHigherOrder,Soda:1995fz,NambuTaruya,
Sasaki:1998ug,Shibata:1999zs,Lyth:2004gb,Tanaka:2006zp}. 

In cosmological situations, the gradient expansion approach is
valid on scales greater than the Hubble horizon scales, and
an advantage of the approach is that we can calculate
perturbations to full nonlinear order in their amplitudes.
This is particularly useful when dealing with the general
non-Gaussianity for which it may be necessary 
to evaluate not only second-order perturbative corrections but also
higher order corrections.

At the leading order in the gradient expansion, i.e. neglecting all
the spatial gradients, Lyth, Malik and Sasaki~\cite{Lyth:2004gb} 
studied nonlinear scalar curvature perturbations, proved the nonlinear 
$\delta N$ formula, and constructed a gauge invariant (time-slice
independent) nonlinear scalar curvature perturbation.
Although this leading order approximation is sufficient
for a large class of inflation models,
there are models for which it is necessary to take into
account the next order corrections. For example, in the
context of the standard linear perturbation theory, Leach et al.
 pointed out that there can be enhancement of
the comoving curvature perturbation on superhorizon scales,
where it is usually conserved, even in a single field model
if the slow-roll condition is temporarily violated~\cite{Leach:2001zf}.
There it was shown that $O(k^2)$ corrections to the curvature perturbation
on superhorizon, where $k$ is the comoving wavenumber,
play a crucial role for the enhancement.
Because $O(k^2)$ corrections correspond to $O(\epsilon^2)$ terms
in gradient expansion, this implies that it is necessary to
include $O(\epsilon^2)$ terms in such a model. 
Then, we expect that the enhancement may give rise to large
non-Gaussianity. Indeed,
Chen et al.~\cite{Chen:2006xj} numerically found in a single-field
inflation that large non-Gaussianity can be generated if the slow-roll
condition is temporarily violated, using third-order action derived by
Maldacena~\cite{Maldacena:2002vr}.

In this paper, focusing on a single-field inflation,
we formulate the gradient expansion to $O(\epsilon^2)$
on the uniform Hubble slicing. 
In most of the previous studies, either only the leading order
terms in gradient expansion was discussed or the choice of
time-slicing was not quite adequate for the study of 
non-Gaussianity from inflation. 
Here we adopt the uniform Hubble slicing because the curvature
perturbation on this slicing directly determines
the initial condition for the CMB anisotropies and
the large scale structure of the universe.

We employ the $(3+1)$-decomposition of the Einstein equations, and
consider a single scalar field with an arbitrary potential.
We then derive the general solution for all the variables.
As discussed in the case of a perfect fluid in~\cite{Tanaka:2006zp},
 we find that the identification of the tensor mode in the
spatial metric is rather arbitrary, depending on how one
fixes the spatial coordinates, as a reflection of general
covariance, while it can be unambiguously identified in
the extrinsic curvature of the metric.

This paper is organized as follows. In Section~\ref{sec:gradexpand},
we define basic variables, and describe the assumptions
we adopt in gradient expansion.
Then we present the general solution for all the physical quantities
to $O(\epsilon^2)$ on the uniform Hubble slicing. 
In Section~\ref{sec:proof},
we discuss the validity of the assumption adopted in
Section~\ref{sec:gradexpand} by appealing to linear theory and
by considering the vacuum fluctuations of the scalar
field at and around horizon crossing. We conclude the paper in Section IV.
In Appendix~\ref{sec:basic}, the basic equations are presented. 
In Appendix~\ref{sec:order}, the estimation of the orders
of physical quantities in powers of $\epsilon$ is given.
In Appendix~\ref{sec:derivation}, the general solutions for all
 the physical quantities are derived.

%%%%%%%%%%%%%%%%%%%%%%%%%%%%%%%
%%%  Sec. II   subsection A  %%
%%%%%%%%%%%%%%%%%%%%%%%%%%%%%%%
\section{Gradient expansion}
\label{sec:gradexpand}

\subsection{Basic variables}

In the $(3+1)$-decomposition, the metric is expressed as
\begin{eqnarray}
ds^2&=&g_{\mu\nu}dx^{\mu}dx^{\nu}
\nonumber \\
&=&(-\alpha^2+\beta_k\beta^k)dt^2+2\beta_idx^idt+\gamma_{ij}dx^idx^j,
\label{eq:metric}
\end{eqnarray}
where $\alpha$, $\beta^i$ ($\beta^i=\gamma^{ij}\beta_j)$, 
and $\gamma_{ij}$ are the lapse function, shift vector, and 
the spatial metric, respectively. We rewrite $\gamma_{ij}$ as
\begin{eqnarray}
\gamma_{ij}(t,x^k)=a^2(t)\,\psi^4(t,x^k)\,\tilde{\gamma}_{ij}(t,x^k)\,;
\quad \det(\tilde{\gamma}_{ij})=1\,,
\label{eq:3metric}
\end{eqnarray}
where the function $a(t)$ is the scale factor of a fiducial
homogeneous and isotropic background universe.

The extrinsic curvature $K_{ij}$ is defined by
\begin{eqnarray}
K_{ij}\equiv -\nabla_in_{j}\,,
\label{Kdef}
\end{eqnarray}
where $n_\mu=(-\alpha,0,0,0)$ is the vector unit normal 
to the time slices. We decompose the extrinsic curvature as
\begin{equation}
K_{ij}=\frac{\gamma_{ij}}{3}K+\psi^4a^2\tilde{A}_{ij}\,;
\quad K\equiv \gamma^{ij}K_{ij}\,,
\label{eq:Kij}
\end{equation}
where $\tilde{A}_{ij}$ represents the traceless part of $K_{ij}$. 
The indices of $K_{ij}$ are to be raised or lowered by $\gamma^{ij}$
 and $\gamma_{ij}$, and the indices of $\tilde{A}_{ij}$ by 
$\tilde{\gamma}^{ij}$ and $\tilde{\gamma}_{ij}$.

 The stress-energy tensor for a single scalar field is 
\begin{equation}
T_{\mu\nu}=\nabla_{\mu}\phi\nabla_{\nu}\phi-\frac{1}{2}g_{\mu\nu}(\nabla^{\alpha}\phi\nabla_{\alpha}\phi+2V(\phi)).
\label{eq:emtensor}
\end{equation}
We define the local Hubble parameter as $1/3$ of the expansion of
the unit normal vector $n^\mu$, which is
equal to $-1/3$ of the trace of the extrinsic curvature
in our convention,
\begin{equation}
3H\equiv -K=\frac{3\dot{a}}{\alpha a}+6\frac{\partial_t{\psi}}{\alpha\psi}
-\frac{D_i \beta^i}{\alpha} \,,
\label{eq:Hubbledef}
\end{equation}
where a dot $\dot{~}$ denotes $d/dt$ and
$D_i$ is the covariant derivative with respect to $\gamma_{ij}$.
In the following, we adopt the uniform Hubble slicing.
For this slicing, we have
\begin{equation}
H(t)=\frac{\dot{a}}{a}\,,
\label{eq:uniHubble}
\end{equation}
and Eq.~(\ref{eq:Hubbledef}) implies
\begin{equation}
\alpha
=\displaystyle1+\frac{2\partial_t{\psi}}{H\psi}-\frac{D_i\beta^i}{3H}
\,.
\label{eq:alpha}
\end{equation}

%%%%%%%%%%%%%%%%%%%%%%%%%%%%%%%%%%%
% Sec.2 subsec.2 Gradient expansion scheme
%%%%%%%%%%%%%%%%%%%%%%%%%%%%%%%%%%%
\subsection{Expansion scheme}

We investigate nonlinear superhorizon perturbations with the gradient
expansion approach, which is called by various names by various authors,
the quasi-isotropic 
expansion~\cite{Lifshitz:1963ps,Belinsky:1982pk,Muller:1989rp,Khalatnikov:2003ac},
the anti-Newtonian approximation \cite{antiNewton,Tomita:1975kj},
the spatial gradient expansion~\cite{Salopek:1990jq,Comer:1994np,Deruelle:1994iz,
Iguchi,Salopek:1990mp,HJHigherOrder,Soda:1995fz,NambuTaruya}, or 
the long wavelength approximation \cite{Sasaki:1998ug,Shibata:1999zs}.
 In this approach, we assume that the characteristic
 length scale $L$ of inhomogeneities is always much larger than the
 Hubble horizon scale, $L\gg H^{-1}\sim t$. We introduce a small parameter
 $\epsilon$, and assume that $L$ is of $O(1/\epsilon)$. 
This assumption is equivalent to assuming that the magnitude of spatial
gradients is given by $\partial_i \psi=\psi\times O(\epsilon)$,
$\partial_i \alpha=\alpha\times O(\epsilon)$, etc.. In the limit
 $L\rightarrow\infty$, i.e., $\epsilon\rightarrow 0$, the universe
looks locally like a FLRW spacetime, where 'locally' means as seen
on the scale of the Hubble horizon volume. 
It is noted that physical quantities which are approximately
homogeneous on each Hubble horizon scale can vary nonlinearly
on very large scales.

The local homogeneity and isotropy imply that $\beta^i=O(\epsilon)$ and
 $\partial_t{\tilde{\gamma}}_{ij}=O(\epsilon)$ because the local FLRW 
equations should be realized in the limit $\epsilon\rightarrow 0$. 
However, we further assume that
$\beta^i=O(\epsilon^3)$ and $\partial_t{\tilde{\gamma}}_{ij}=O(\epsilon^2)$.
Technically, these additional assumptions make the analysis
of $O(\epsilon^2)$ correction terms much simpler,
as discussed in Appendix~\ref{sec:derivation}.1.
Physically, of course, it is necessary to justify them. 
For the former assumption on $\beta^i$, since it is just 
a matter of choice of the spatial coordinates,
this does not cause any loss of generality.
In fact, once we obtain the solution, it is straightforward
to express it in a more general spatial coordinate system
by the coordinate transformation $x^i\to\bar{x}^i=F^i(t,x^k)$
such that $\partial_tF^i=O(\epsilon)$, corresponding
to $\bar\beta^i=O(\epsilon)$ in the new coordinate system.
As for the latter assumption on $\partial_t{\tilde{\gamma}}_{ij}$,
however, it is not simply a matter of choice.
In Sec.~\ref{sec:proof}, with the help of the result from
linear theory, we give a convincing argument, if not rigorous,
that this assumption is indeed satisfied for perturbations
arising from the vacuum fluctuations.
To summarize, our basic assumptions are
\begin{eqnarray}
\beta^i=O(\epsilon^3)\,,
\quad
\partial_t{\tilde{\gamma}}_{ij}=O(\epsilon^2)\,.
\label{eq:assumptions}
\end{eqnarray}
Applying these assumptions to the Einstein-scalar field equations,
we find
\begin{eqnarray}
&&
\psi=O(1)\,,
\quad \alpha-1=O(\epsilon^2)\,,
\quad \partial_t{\psi}=O(\epsilon^2)\,,
\quad \tilde{A}_{ij}=O(\epsilon^2)\,.
\label{eq:ordercount}
\end{eqnarray}
These estimates are derived in Appendix~\ref{sec:order}.
Here one comment is in order. The fact that $\partial_t{\psi}=O(\epsilon^2)$
means $\psi$ is conserved if the $O(\epsilon^2)$ corrections can be neglected.
This result was derived by Salopek and Bond~\cite{Salopek:1990jq} 
for a single scalar field system, and 
by Lyth, Malik, and Sasaki~\cite{Lyth:2004gb} for more general systems.

%%%%%%%%%%%%%%%%%%%%%%%%%%%%%%%%%%%%
%% Sec.2 Sub.C the general solutions
%%%%%%%%%%%%%%%%%%%%%%%%%%%%%%%%%%%%
\subsection{General solution}

Here we present the general solution for all the physical quantities,
valid to $O(\epsilon^2)$ in gradient expansion.
We defer the derivation to Appendix~\ref{sec:derivation} because
it is not much different from the one we gave in the previous
paper~\cite{Tanaka:2006zp}, except for the fact that the present
paper deals with the case of a scalar field while the previous paper
dealt with a perfect fluid.

The general solution is
\begin{eqnarray}
\alpha&=&1+2\frac{\dot\phi_{\ast}}{a^3\dot\phi^3}
\left[
\pret C(x^k)\left(2a\,\dot\phi+\frac{dV}{d\phi}\int^t_{t_*}a(t')dt'\right)
+\pret D(x^k)\frac{dV}{d\phi}\right],
\label{eq:summary1} \\
\psi&=&\prez L(x^k)\Bigl(1+\frac{1}{2}\int^t_{t_{\ast}} (\alpha-1)Hdt'\Bigr),
\label{eq:summary2} \\
\tilde \gamma_{ij}
&=&\prez f_{ik}(x^\ell)\Bigl(\delta^k_j-2\pret F^k{}_j(x^\ell)\int^t_{t_{\ast}}  
\frac{dt^{\prime}}{a^3(t^{\prime})}\int^{t^{\prime}}_{t_{\ast}}  
a(t^{\prime\prime})dt^{\prime\prime}
-2\pret C^k{}_j(x^\ell)\int^t_{t_{\ast}} \frac{dt^{\prime}}{a^3(t^{\prime})}\Bigr),
\label{eq:summary3} \\
\tilde{A}_{ij}&=&\frac{\pret F_{ij}(x^k)}{a^3}\int^t_{t_{\ast}} 
 a(t^{\prime})dt^{\prime}+\frac{\pret C_{ij}(x^k)}{a^3},
\label{eq:summary4} \\
\tilde\phi&=&\phi(t)
+\pret C(x^k)\frac{\dot\phi_*}{a^3\dot\phi}\int^t_{t_*} a(t')dt'
+\pret D(x^k)\frac{\dot\phi_*}{a^3\dot\phi}\,,
\label{eq:summary5} 
\end{eqnarray}
where the index $(n)$ is attached to a quantity of $O(\epsilon^n)$
except for the scalar field. For notational simplicity, the full
scalar field is denoted by $\tilde\phi$, and the lowest order scalar field
$\prez\phi$, which depends only on time, is denoted simply by $\phi$.
The time $t_{\ast}$ is an arbitrary reference time and $\dot\phi_*=\dot\phi(t_*)$.
The tensor $\pret F_{ij}$ and the scalar $\pret C$ are given 
by Eqs.~(\ref{eq:Fijdef}) and (\ref{eq:Cdef}), respectively, as functions of
$\prez L$ and $\prez f_{ij}$. 
The tensor $\pret C_{ij}$ is traceless with respect to $\prez f_{ij}$,
The function $\pret D$ is related to $\pret C_{ij}$ through
the momentum constraint~(\ref{eq:CC}). 

To clarify the physical role of these freely specifiable functions,
let us first count the degrees of freedom. Since 
$\pret C$ and $\pret F_{ij}$ are determined by 
$\prez L$ and $\prez f_{ij}$, they have no degree of freedom.
Since the determinant of
$\prez f_{ij}$ is unity, it has 5 degrees of freedom,
and since $\pret C_{ij}$ is traceless, it also has 5 degrees of freedom.
In addition we have $\prez L$ and $\pret D$, each of which
counts 1 degree of freedom. The momentum constraints which consist
of 3 equations, relate $\pret C_{ij}$ to $\pret D$, and reduce the total
degrees of freedom by 3. 
So, adding together, the total number is $5+5+1+1-3=9$,
while the true physical degrees of freedom
are $2+2+2=6$, where $2$ are of the single scalar field (1 for a growing mode
 and 1 for a decaying mode) and $2+2$ are of the gravitational waves
 (2 for the metric and 2 for the extrinsic curvature). This implies
that there still remains 3 degrees of freedom. 

As discussed in the case of a single perfect fluid in the previous
paper~\cite{Tanaka:2006zp}, the remaining 3 degrees of freedom 
come from the spatial covariance, that is, from the gauge freedom of
purely spatial coordinate transformations
$x^i\to \bar x^i=f^i(x^j)$. Thus we may regard that
$\prez f_{ij}$ contains these 3 gauge degrees of freedom.
The correspondence between the nonlinear solutions and the solutions in linear
 theory can be understood from the time dependence of the nonlinear solutions.
 So, we see that $\prez L$ and $\prez f_{ij}$ represent growing modes,
 and $\pret D$ and $\pret C_{ij}$ represent decaying modes.

To summarize, the degrees of freedom contained in the
freely specifiable functions can be interpreted as
\begin{eqnarray}
\prez L~ &\cdots&\ 1=1~\mbox{(scalar growing mode)}\,,
\nonumber\\
\prez f_{ij}~&\cdots&\ 5=2~\mbox{(tensor growing modes)}+3~\mbox{(gauge modes)}\,,
\nonumber\\
\pret C_{ij}~&\cdots&\ 5=2~\mbox{(tensor decaying modes)}+3~\mbox{(constraints)}\,,
\nonumber \\
\pret D~ &\cdots&\ 1=1~\mbox{(scalar decaying mode)}\,.
\label{eq:dof}
\end{eqnarray}
To fix the gauge completely, one has to impose 
3 spatial gauge conditions on $\prez f_{ij}$ to extract the
physical tensor degrees of freedom. As discussed 
in Ref.~\cite{Tanaka:2006zp}, this cannot be done in a spatially
covariant way because $\prez f_{ij}$ is the metric and any covariant derivative
of it vanishes identically. Thus the tensor modes cannot be extracted out
from $\prez f_{ij}$ unless we introduce a certain 'background' metric.
This is an important difference from the linear case in which there exists
a background metric.

In contrast, as for the extrinsic curvature, we may identify the tensor 
modes in it, because its transverse-traceless part can be extracted
unambiguously~\cite{York:1973ia,Tanaka:2006zp}.
Namely, the transverse part of $\prez L^6\tilde A_{ij}$ can be determined
uniquely and it can be identified as the tensor modes.
As noted in Ref.~\cite{Tanaka:2006zp}, this implies that the tensor
modes in the extrinsic curvature are determined non-locally,
and they exist even for a trivial $\prez f_{ij}$, say 
for $\prez f_{ij}=\delta_{ij}$. This generation of tensor modes in 
the extrinsic curvature is a result of nonlinear interactions of the
scalar modes. We note, however, that it is not obvious if the tensor
modes we identified can be called gravitational waves. To make a clear
connection between the tensor modes and gravitational
waves, it is necessary to evolve the system until
the scale of interest becomes sufficiently smaller
than the Hubble scale. But this is beyond the scope
of the present paper.

%%%%%%%%%%%%%%%%%%%%%%%%%%%%%%%%%%%%%%%%%%%
%%Sec.3 
%%%%%%%%%%%%%%%%%%%%%%%%%%%%%%%%%%%%%%%%%5
\section{validity of the assumption in the linear limit}
\label{sec:proof}

In this section, we discuss the validity of our central assumption,
$\partial_t{\tilde\gamma}_{ij}=O(\epsilon^2)$. With the help of
linear theory, we argue that it can be physically justified.
We first consider the equation for the curvature perturbation
during inflation, and argue that
the condition $\partial_t{\tilde\gamma}_{ij}=O(\epsilon^2)$
 is naturally satisfied.
Then considering the quantum fluctuations as the source of the
curvature perturbation, we explicitly show that this assumption
holds not only in the case of slow-roll inflation
but also in the case of the Starobinsky model, in which
the slow-roll condition is temporarily violated due to
a sudden change in the slope of the potential.

Since gradient expansion is effective only on superhorizon scales,
as the initial condition for the quantities we calculate,
we need to know their behavior when their wavelength
crosses the Hubble horizon radius. To do so, we assume that the
linear perturbation is a sufficiently good approximation up to and around
the Hubble horizon scale. For linearized quantities, we follow the 
notation of Kodama and Sasaki~\cite{Kodama:1985bj}.

As usual, we expand the linearized quantities in terms of
spatial harmonics. The background is a spatially flat FLRW universe.
We introduce the scalar, vector and tensor harmonics,
$Y_{\bm k}$, $Y_{{\bm k}\,i}^{(1)}$ and $Y_{{\bm k}\,ij}^{(2)}$, respectively,
\begin{eqnarray}
&&(\Delta+k^2)Y_{\bm k}=0\,,
\cr
&&(\Delta+k^2)Y_{{\bm k}\,i}^{(1)}=0\,;
\quad \partial^iY_{{\bm k}\,i}^{(1)}=0\,,
\cr
&&(\Delta+k^2)Y_{{\bm k}\,ij}^{(2)}=0\,;
\quad
\delta^{ij}Y_{{\bm k}\,ij}=\partial^iY_{{\bm k}\,ij}^{(2)}=0\,,
\end{eqnarray}
where $\Delta=\partial^i\partial_i$ is the 3-dimensional Laplacian.
For the sake of notational simplicity, we suppress the mode
index ${\bm k}$ below. For a universe dominated by a scalar field,
it is known that the vector perturbations are not be excited.
Therefore, we ignore the vector modes.

We express the metric as
\begin{eqnarray}
ds^2=a(\eta)^2\left[-(1+2AY)d\eta^2
-2BY_jdx^jd\eta+(\delta_{ij}+2H_LY\delta_{ij}
+2H_TY_{ij}+2H_T^{(2)}Y_{ij}^{(2)})dx^idx^j\right],
\label{eq:linearMetric}
\end{eqnarray}
where $d\eta=dt/a(t)$ is the conformal time, and
\begin{eqnarray}
Y_{j}\equiv -k^{-1}\partial_jY\,,
\quad
Y_{ij}\equiv k^{-2}
\left[\partial_i\partial_j-\frac{1}{3}\delta_{ij}\Delta\right]Y\,.
\end{eqnarray}
The correspondences of these to the metric components defined in
Section~\ref{sec:gradexpand} are
\begin{eqnarray}
\alpha=1+AY\,,
\quad
\beta_j=-aBY_j\,,
\quad
\psi^2=1+H_LY\,,
\quad
\tilde\gamma_{ij}=\delta_{ij}+2H_TY_{ij}+2H_T^{(2)}Y_{ij}^{(2)}\,.
\end{eqnarray}
In passing, we note that our assumption $\partial_t\tilde\gamma_{ij}
=O(\epsilon^2)$ corresponds to $\partial_t H_T=O(\epsilon^2)$ and $\partial_t H_T^{(2)}=O(\epsilon^2)$ in the linear limit.
We also introduce the quantities
\begin{eqnarray}
{\cal R}&\equiv& H_L+\frac{H_T}{3}\,,
\label{eq:calRDef}
\\
\sigma_g&\equiv& k^{-1}H_T^{\prime}-B\,,
\label{eq:sigmaDef}
\\
{\cal K}&\equiv& -A+\frac{k}{3{\cal H}}B+{\cal H}^{-1}H_L^{\prime}
=-A+{\cal H}^{-1}{\cal R}'-\frac{k}{3{\cal H}}\sigma_g\,,
\label{eq:calKDef}
\end{eqnarray}
where a prime (${~}'$) denotes a conformal time derivative
$d/d\eta$. These quantities are known to be independent of the choice of
the spatial coordinates but depend only on the choice of time-slicing.
The first one, ${\cal R}$ is called the curvature perturbation
because the spatial curvature scalar $R[\gamma]$ is given by
\begin{eqnarray}
R[\gamma]=-\frac{4}{a^2}\Delta\left[{\cal R}Y\right]\,.
\end{eqnarray}
The variables ${\cal K}$ and $\sigma_g$ represent
the perturbations in the extrinsic curvature,
\begin{eqnarray}
K&=&-3H\left(1+{\cal K}Y\right)\,,
\quad
\tilde A_{ij}=-\frac{k}{a}\sigma_gY_{ij}\,.
\end{eqnarray}

Under the gauge transformation induced by an
infinitesimal change of time-slicing $\bar\eta=\eta+TY$,
${\cal R}$, $\sigma_g$ and ${\cal K}$ transform as
\begin{eqnarray}
\bar{\cal R}&=&{\cal R}-{\cal H}T\,,
\label{eq:RT}
\\
\bar{\sigma}_g&=&\sigma_g-kT\,,
\label{eq:sigmaT}
\\
\bar{\cal K}&=&{\cal K}
+{\cal H}^{-1}\left({\cal H}^2-{\cal H}^{\prime}+\frac{k^2}{3}\right)T\,,
\label{eq:KT}
\end{eqnarray}
where ${\cal H}\equiv aH$ is the conformal Hubble parameter.

Since the standard linear calculation gives the curvature
perturbation on the comoving slicing, we need to relate the quantities
on the comoving slicing to those on the uniform Hubble slicing. 
In particular, analytic expressions for the comoving curvature perturbations
in the Starobinsky model resulting from the quantum fluctuations 
are given in~\cite{Starobinsky:1992ts}.
Here and in what follows, we denote a quantity on the comoving 
and uniform Hubble slicing by the subscripts $c$ and $H$, respectively.
Thus, we first express the geometrical quantities on the comoving slicing
in terms of ${\cal R}_c$. Then we consider an infinitesimal transformation
from the comoving slicing to the uniform Hubble slicing.

The $(0,\mu)$-components of the Einstein equations on the comoving slicing
give
\begin{eqnarray}
\delta G^0_0&=&\kappa^2\delta T^0_0\,;
\nonumber \\
& &
2[3{\cal H}^2A_c+{\cal H}k(\sigma_g)_c
-3{\cal H}{\cal R}_c^{\prime}-k^2{\cal R}_c]=\kappa^2\phi^{\prime 2}A_c\,,
\label{eq:deltaG00}\\
\delta G^0_j&=&\kappa^2\delta T^0_j\,;
\nonumber \\
& &{\cal H}A_c-{\cal R}_c^{\prime}=0\,,
\label{eq:deltaG0j}
\end{eqnarray}
where $\kappa^2=8\pi G$. From these, we have
\begin{eqnarray}
A_c=\frac{1}{{\cal H}}{\cal R}_c'\,,
\quad
k(\sigma_g)_c
=\frac{1}{\cal H}\,
\left(k^2{\cal R}_c
+\frac{\kappa^2}{2}\frac{\phi'{}^2}{{\cal H}}{\cal R}_c'\right)\,,
\end{eqnarray}
where it is noted that the background equation gives the relation,
\begin{eqnarray}
-a^2\dot H=-a\left(\frac{{\cal H}}{a}\right)'=
{\cal H}^2-{\cal H}'=\frac{\kappa^2}{2}\phi'{}^2\,.
\end{eqnarray}
Then the second equality of Eq.~(\ref{eq:calKDef}) gives
\begin{eqnarray}
{\cal K}_c=-\frac{k}{3{\cal H}}(\sigma_g)_c
=-\frac{1}{3{\cal H}^2}
\left(k^2{\cal R}_c+
\frac{\kappa^2\phi^{\prime 2}}{2{\cal H}}{\cal R}_c^{\prime}\right)\,.
\label{eq:calKc}
\end{eqnarray}
Since ${\cal K}$ is zero on the uniform Hubble slicing by definition,
we have
\begin{eqnarray}
0={\cal K}_H={\cal K}_c
+{\cal H}^{-1}({\cal H}^2-{\cal H}^{\prime}+\frac{k^2}{3})T\,.
\label{eq:KuKc}
\end{eqnarray}
This gives $T$ in terms of ${\cal K}_c$, which in turn is given
by Eq.~(\ref{eq:calKc}). The result is
\begin{eqnarray}
{\cal H}T=\frac{1}{3\frac{\kappa^2}{2}\phi^{\prime}{}^2+k^2}
\left(k^2{\cal R}_c+
\frac{\kappa^2\phi^{\prime 2}}{2{\cal H}}{\cal R}_c^{\prime}\right)\,.
\label{eq:T}
\end{eqnarray}
Thus, we obtain
\begin{eqnarray}
{\cal R}_H&=&\left(H_L-\frac{H_T}{3}\right)_H
=
{\cal R}_c-\frac{1}{3\frac{\kappa^2}{2}\phi^{\prime}{}^2+k^2}
\left(k^2{\cal R}_c+
\frac{\kappa^2\phi^{\prime 2}}{2{\cal H}}{\cal R}_c^{\prime}\right)\,,
\label{eq:calRH}
\\
(k\sigma_g)_H&=&(H_T^{\prime}-kB)_H
=\frac{1}{\cal H}\,
\frac{\frac{\kappa^2}{2}\phi'{}^2}{\frac{\kappa^2}{2}\phi'{}^2+
\frac{k^2}{3}}
\left(k^2{\cal R}_c+
\frac{\kappa^2\phi^{\prime 2}}{2{\cal H}}{\cal R}_c^{\prime}
\right)\,.
\label{eq:HuBuRc}
\end{eqnarray}
Since $B=O(k)$ because $\beta^i=O(\epsilon)$,
the above results imply that our assumption 
$\partial_t\tilde\gamma_{ij}=O(\epsilon^2)$, which corresponds
to $H_T'=O(k^2)H_T$ in the linear limit, is justified
if ${\cal R}_c'=O(k^2){\cal R}_c$
on superhorizon scales.

Now let us argue that this is indeed so in general.
We know that ${\cal R}_c$ satisfies the equation,
\begin{eqnarray}
{\cal R}_c^{\prime\prime}
+2\frac{z^{\prime}}{z}{\cal R}_c^{\prime}+k^2{\cal R}_c=0\,;
\quad z\equiv \frac{a\phi'}{{\cal H}}=\frac{a\dot\phi}{H}\,.
\label{eq:Rceq}
\end{eqnarray}
We consider an inflationary universe in which we approximately
have ${\cal H}\sim -1/\eta$. 
There are two independent solutions $u$ and $v$, which are
functions of $k\eta$. We may fix these solutions such that
they behave in the superhorizon limit $k/{\cal H}\sim k|\eta|\to0$ as
\begin{eqnarray}
u&=&\left[1+O(k^2)\right]u_0\,;\quad u_0=\mbox{const.}\,,
\cr
v&=&\left[1+O(k^2)\right]v_0\,;\quad
 v_0\equiv
\frac{\displaystyle\int_\eta^0d\eta'/z^{2}(\eta')}
{\displaystyle\int_{\eta_k}^0d\eta'/z^{2}(\eta')}\,,
\end{eqnarray}
where $\eta_k\sim -k^{-1}$ is the horizon crossing time,
 ${\cal H}(\eta_k)\equiv k$, and we have used the fact
that the long wavelength expansion of 
Eq.~(\ref{eq:Rceq}) would give corrections only in powers of $k^2$.
Then the general solution for ${\cal R}_c$ is given by
\begin{eqnarray}
{\cal R}_c=Au+Bv\,,
\end{eqnarray}
where $A$ and $B$ are constants of order unity in the
sense of the $k$ expansion.
Thus at $k|\eta|\ll1$, we have
\begin{eqnarray}
{\cal R}_c=Au_0\,,\quad
{\cal R}_c'=O(k^2)Au_0+Bv_0'\,.
\end{eqnarray}
Now, assuming that the violation of the slow-roll condition 
is not too strong, which is necessary for the spectrum to
be approximately scale-invariant, we have $z\propto a\propto\eta^{-1}$,
hence
\begin{eqnarray}
v_0'=-\left(\int_{\eta_k}^0d\eta'\frac{z^2(\eta)}{z^{2}(\eta')}\right)^{-1}
=O(k^3)\,.
\end{eqnarray}
Thus we conclude that ${\cal R}_c'=O(k^2){\cal R}_c$
on superhorizon scales.

To reinforce the above argument, let us consider the Starobinsky
model in which the slow-roll condition can be significantly
violated~\cite{Starobinsky:1992ts}. The Starobinsky model 
has a sudden change in its slope at $\phi=\phi_0$ such that
\begin{eqnarray}
V(\phi)=
\begin{cases}
V_0+A_+(\phi-\phi_0)\qquad \mbox{for}\quad \phi > \phi_0\,, 
\\
V_0+A_-(\phi-\phi_0)\qquad \mbox{for}\quad \phi < \phi_0\,,
\end{cases}
\end{eqnarray}
where $A_+$, $A_-$ and $\phi_0$ are assumed to be positive so that
the scalar field evolves from a large positive value of
$\phi$ toward $\phi=0$.
Then the background scalar field $\phi$ satisfies
\begin{eqnarray}
3H\dot\phi=
\left\{\begin{array}{ll}
-A_+\qquad & \mbox{for}\quad  \phi > \phi_0\,,
\\
-\left(A_-+(A_+-A_-)e^{-3H(t-t_0)}\right)\qquad
 & \mbox{for} \quad  \phi < \phi_0\,,
\end{array}\right.
\label{eq:dotphiSta}
\end{eqnarray}
where $t_0$ is the time at which $\phi=\phi_0$,
and the de Sitter approximation, $3H^2=\kappa^2V_0$ 
is assumed to be valid. 
Thus the scalar field slow-rolls at
 $\phi>\phi_0$, and violates the slow-roll condition temporarily at
$\phi<\phi_0$. The evolution is decelerated
if ${A_+}/{A_-}>1$, or accelerated if ${A_+}/{A_-}<1$,
compared to the slow-roll evolution.

The curvature perturbation on the comoving slicing during
the non-slow-roll regime at $\phi\le\phi_0$ is~\cite{Starobinsky:1992ts}
\begin{eqnarray}
{\cal R}_c&=&-\frac{iH^2}{\sqrt{2}k^{3/2}\dot\phi}\Bigl(\alpha(k)
e^{-ik\eta}(1+ik\eta)-\beta(k)e^{ik\eta}(1-ik\eta)\Bigr)\,,
\label{eq:Rcnonslow}
\\
& &\alpha(k)\equiv 1+\frac{3i}{2}\left(\frac{A_-}{A_+}-1\right)\frac{k_0}{k}
\left(1+\frac{k^2_0}{k^2}\right),
\label{eq:alpha(k)}
\\
& &\beta(k)\equiv-\frac{3i}{2}\left(\frac{A_-}{A_+}-1\right)
e^{2i{k}/{k_0}}\frac{k_0}{k}\left(1+\frac{ik_0}{k}\right)^2,
\label{eq:beta(k)}\\
& &\quad |\alpha|^2-|\beta|^2=1,
\nonumber
\end{eqnarray}
where $k_0=(a(t_0)H_0)^{-1}$. 
On superhorizon scales, we have $k|\eta|\ll 1$.
The standard slow-roll case corresponds to $A_+=A_-$. In this case,
we immediately see that $\alpha(k)=1$, $\beta(k)=0$, and
${\cal R}_c'=O(k^2){\cal R}_c$ on superhorizon scales.

We now turn to the case $A_+\neq A_-$. We assume $A_-/A_+=O(1)$
so that the slow-roll condition is not severely violated. For $k<k_0$,
we expand the three exponentials in Eq.~(\ref{eq:Rcnonslow}) to obtain
\begin{eqnarray}
{\cal R}_c=\frac{3iH^3}{\sqrt{2}k^{3/2}A_+}
\left(1-\frac{A_+}{3H\dot\phi}\frac{(k\eta)^2}{2}
\Bigl[1+\Bigl(\frac{A_-}{A_+}-1\Bigr)
\Bigl(1+2k_0\eta-\frac{(k_0\eta)^3}{5}
-\frac{4}{5(k_0\eta)^2}\Bigr)+O(k)
\Bigr]\right).
\label{eq:RcTok2}
\end{eqnarray}
Thus we have ${\cal R}_c'=O(k^2){\cal R}_c$.
For $k>k_0$, we have $\alpha(k)=1+O(k_0/k)$ and
$\beta(k)=O(k_0/k)$. Hence
\begin{eqnarray}
{\cal R}_c'
=\left[-a\frac{\ddot\phi}{\dot\phi}+O(k^2)\right]{\cal R}_c
=\left[O\left((k\eta)^2\right)
\left(\frac{k_0}{k}\right)^2+O(k^2)\right]{\cal R}_c\,.
\end{eqnarray}
Since $k>k_0$, we see the right-hand side of this is also 
of $O(k^2){\cal R}_c$. Thus we have shown that
${\cal R}_c'=O(k^2){\cal R}_c$ holds even for the Starobinsky model
in which the slow-roll condition can be violated. This in turn
implies $H_T'=O(k^2)H_T$. 

Next, we consider the tensor perturbation $H_T^{(2)}$. The evolution
equation is
\begin{eqnarray}
H_T^{(2)}{}''+2\frac{a'}{a}H_T^{(2)}{}'+k^2H_T^{(2)}=0\,.
\label{eq:HT2eq}
\end{eqnarray}
This equation has the same structure as Eq.~(\ref{eq:Rceq})
for ${\cal R}_c$ if we replace $z$ with $a$. Then we can
repeat exactly the same argument and conclude that
$H_T^{(2)}{}'=O(k^2)H_T^{(2)}$.

Again, this conclusion
can be supported by considering the quantum fluctuations
explicitly. When the de Sitter approximation $H=$const. holds,
the tensor perturbation from the quantum fluctuations 
during inflation are given by
\begin{eqnarray}
H_T^{(2)}=\frac{\sqrt{2\pi}H}{k^{3/2}}(k\eta-i)e^{-ik\eta}.
\label{eq:HT2}
\end{eqnarray}
Taking the conformal time derivative of this equation, 
we obtain at $k|\eta|\ll1$,
\begin{eqnarray}
H_T^{(2)\prime}=-\frac{\sqrt{2\pi}H}{k^{3/2}}(ik^2\eta)e^{-ik\eta}
=O(k^2) H_T^{(2)}.
\label{eq:HT2prime}
\end{eqnarray}

Thus, provided that the linear theory is a good approximation
up to scales corresponding to a few e-foldings after horizon crossing,
our assumption $\partial_t{\tilde\gamma}_{ij}=O(\epsilon^2)$ is
justfied for perturbations originated from the quantum fluctuations
inside the horizon.

%%%%%%%%%%%%%%%%%%%%%%%%%%%%%%%%%%%%
%% Sec. Conclusion 
%%%%%%%%%%%%%%%%%%%%%%%%%%%%%%%%%%%%

\section{Conclusion}
\label{sec:conclusion}

In this paper, taking the gradient expansion approach,
we have investigated nonlinear perturbations on 
superhorizon scales in a universe dominated by a single scalar field.
 We have derived the general solution for all the physical quantities
valid to second order in the spatial gradients on the uniform Hubble
 slicing. In particular, an expression for the nonlinear curvature 
perturbation, which plays a central role in the evaluation of a possible
non-Gaussianity from inflation, has been obtained.

Parallel to our previous paper~\cite{Tanaka:2006zp}, we have identified
the tensor modes in the extrinsic curvature, while the identification of
the tensor modes in the metric is arbitrary unless we specify the 
background spatial metric. This is an important difference from
the case of linear theory in which the background metric is
uniquely given.

In our analysis, we have adopted a non-trivial assumption
 $\partial_t{\tilde\gamma}_{ij}=O(\epsilon^2)$, which cannot be
justified within the context of gradient expansion.
To justify it, we have appealed to linear theory, and argued that
the assumption $\partial_t{\tilde\gamma}_{ij}=O(\epsilon^2)$
is naturally satisfied in the linear limit.
In particular, as an explicit example, we have considered the
Starobinsky model, in which the slow-roll condition is temporarily
violated due to a sudden change in the slope of the potential,
and we have explicitly shown that the quantum fluctuations indeed
satisfy the assumption $\partial_t{\tilde\gamma}_{ij}=O(\epsilon^2)$.

As mentioned in Introduction, in models in which the slow-roll
condition is temporarily violated as in the case of the Starobinsky
model, the $O(k^2)$ corrections to the curvature perturbation,
which correspond to $O(\epsilon^2)$ corrections in gradient expansion,
may play a crucial role in the determination of the final amplitude
of the curvature perturbation. Since the result of this
paper is valid to $O(\epsilon^2)$, it provides a very useful tool
to investigate the nonlinear behavior of the curvature perturbation
on superhorizon scales for a wide class of models including those
which violate the slow-roll condition. In particular, matching our
result with the quantum fluctuations on scales slightly beyond the
horizon scale, the non-Gaussianity arising from the nonlinear
dynamics of the scalar field on superhorizon scales can be studied.
Applications to specific models of inflation will be
considered in a future publication.

%%%%%%%%%%%%%%%%%%%%%%%%
\section*{Acknowledgements}

This work was supported in part by 
the Monbu-Kagakusho 21st century COE Program 
``Center for Diversity and Universality in Physics",
and by JSPS Grants-in-Aid for Scientific Research
 (B) No.~17340075, and (A) No.~18204024.

%%%%%%%%%%%%%%%%%%%%%%%%%%%%%%%%%%%%
%% Appendix
%%%%%%%%%%%%%%%%%%%%%%%%%%%%%%%%%%%

\appendix

\section{Basic equations}
\label{sec:basic}
Here we show the basic equations for nonlinear quantities.
%%%%%%%%%%%%%%%%%%%%%%
%% Basic equations  %%
%%%%%%%%%%%%%%%%%%%%%%
The Klein-Gordon equation is 
\begin{eqnarray}
\frac{1}{\sqrt{-g}}\frac{\partial}{\partial x^{\mu}}\Bigl[\sqrt{-g}g^{\mu\nu}
\frac{\partial}{\partial x^{\nu}}
\phi\Bigr]-\frac{dV}{d\phi}=0
\label{eq:KGeq}
\end{eqnarray}
%%%%%%%%%%%%%%%%%%%%%%%%
%% Einstein equations
%%%%%%%%%%%%%%%%%%%%%%%%
We employ the $(3+1)$-formalism of the Einstein equations and then the
dynamical variables are $\gamma_{ij}$ and $K_{ij}$.
The $(n,n)$ and $(n,j)$ components of the Einstein equations
give the Hamiltonian and momentum constraint equations, respectively,
while the $(i,j)$ components gives the evolution equations for $K_{ij}$. 
The evolution equations for $\gamma_{ij}$ are given by
the definitions of the extrinsic curvature~(\ref{Kdef}).

In the present case, the Hamiltonian and momentum constraints
are
\begin{eqnarray}
&&R-\tilde{A}_{ij}\tilde{A}^{ij}+\frac{2}{3}K^2=16\pi GE\,,
\label{eq:hamconst} 
\\
&& D_i \tilde{A}^i{}_j-\frac{2}{3}D_jK=8\pi GJ_j \,,
\label{eq:momconst}
\end{eqnarray}
\begin{eqnarray}
E\equiv T_{\mu \nu}n^{\mu}n^{\nu}\,,
\quad
J_j\equiv -T_{\mu \nu}n^{\mu}\gamma^{\nu}_j\,.
\label{eq:EJdef}
\end{eqnarray}

The evolution equations for $\gamma_{ij}$ are given as
\begin{eqnarray}
&&(\partial_t-\beta^k\partial_k)\psi+\frac{\dot{a}}{2a}\psi
=\frac{\psi}{6}\{-\alpha K+\partial_k\beta^k \},
\label{eq:dotpsi} \\
&&(\partial_t-\beta^k\partial_k)\tilde{\gamma}_{ij}=-2\alpha  \tilde{A}_{ij}
+\tilde{\gamma}_{ik} \partial_j\beta^k+\tilde{\gamma}_{jk}\partial_i\beta^k
-\frac{2}{3}\tilde{\gamma}_{ij}\partial_k\beta^k\,,
\label{eq:dotgamma}
\end{eqnarray}
where $\dot{}=d/dt$.

The evolution equations for $K_{ij}$ are given as
\begin{eqnarray}
(\partial_t-\beta^k\partial_k)K
&=&\alpha(\tilde{A}_{ij}\tilde{A}^{ij}
+\frac{1}{3}K^2)-D_kD^k\alpha+4\pi G\alpha(E+S^k{}_k),
\label{eq:dotK} 
\\
(\partial_t-\beta^k\partial_k)\tilde{A}_{ij}
&=&\frac{1}{a^2\psi^4}
[\alpha(R_{ij} -\frac{\gamma_{ij}}{3}R)
-(D_iD_j\alpha-\frac{\gamma_{ij}}{3}D_kD^k\alpha)]
\nonumber  \\
&&+\alpha(K\tilde{A}_{ij}-2\tilde{A}_{ik}\tilde{A}^k{}_j)
\nonumber\\
&& +\tilde{A}_{ik}\partial_j\beta^k+\tilde{A}_{jk}\partial_i\beta^k
-\frac{2}{3}\tilde{A}_{ij}\partial_k\beta^k
-\frac{8\pi G\alpha}{a^2\psi^4}(S_{ij}-\frac{\gamma_{ij}}{3}S^k{}_k),
\label{eq:dotA} 
\end{eqnarray}
where $R_{ij}$ is the Ricci tensor of the metric $\gamma_{ij}$,
$R\equiv \gamma^{ij}R_{ij}$, $D_i$ is the covariant derivative 
with respect to $\gamma_{ij}$, and
\begin{eqnarray}
S_{ij}\equiv T_{ij}\,,
\quad
S^k{}_k\equiv \gamma^{kl}S_{lk}\,.
\label{eq:Sdef}
\end{eqnarray}

\section{Order estimation}
\label{sec:order}
Here we evaluate the order of magnitude of the basic
variables using the equations presented in 
Appendix~\ref{sec:basic} and the assumptions given by 
Eq.~(\ref{eq:assumptions}), namely
\begin{eqnarray}
\beta^i=O(\epsilon^3)\,,
\quad
\partial_t{\tilde{\gamma}}_{ij}=O(\epsilon^2)\,.
\end{eqnarray}

With these assumptions, Eqs.~(\ref{eq:momconst}) and (\ref{eq:dotgamma}) on 
the uniform Hubble slicing yield
\begin{equation}
J_j=O(\epsilon^3).
\label{eq:J3}
\end{equation}
Using Eq.~(\ref{eq:emtensor}), $J_j$ is express as
\begin{eqnarray}
J_j=-\frac{1}{\alpha}\partial_0\phi\partial_j\phi
+\frac{\beta^l}{\alpha}\partial_l\phi\partial_j\phi\,.
\label{eq:Jphi} 
\end{eqnarray}
We expand $\phi$ as $\phi={}_{(0)}\phi+{}_{(1)}\phi+{}_{(2)}\phi+\cdots$.
Then, to satisfy Eq.(\ref{eq:J3}), ${}_{(0)}\phi$ should depend only on time,
and ${}_{(1)}\phi$ should vanish. Thus, we obtain 
\begin{eqnarray}
\postz\phi={}_{(0)}\phi(t)\,,
\quad {}_{(1)}\phi=0.
\label{eq:phi01}
\end{eqnarray}
For notational simplicity, we denote the full scalar field by $\tilde\phi$
and ${}_{(0)}\phi$ by $\phi$ in the following. Thus
\begin{eqnarray}
\tilde\phi(t,x^i)=\phi(t)+\pret\phi(t,x^i)+\cdots.
\label{eq:expandphi}
\end{eqnarray}

{}From the $O(\epsilon^0)$ part of the Hamiltonian constraint, 
Eq.~(\ref{eq:hamconst}), we have 
\begin{equation}
\frac{1}{3}K^2(t)=3H^2=8\pi G\Bigl(\frac{\dot\phi^2}{2\postz\alpha^2}
+V(\phi)\Bigr),
\label{eq:Hamconst0}
\end{equation}
where we have also expanded $\alpha$ as 
$\postz\alpha+{}_{(2)}\alpha+\cdots$. From this equation,
we find that ${}_{(0)}\alpha$ depend only on time,
\begin{equation}
{}_{(0)}\alpha={}_{(0)}\alpha(t).
\label{eq:alpha0}
\end{equation}
Since ${}_{(0)}\alpha$ is spatial homogeneous, 
we may choose the time coordinate to set it to unity, ${}_{(0)}\alpha=1$.
Thus, from Eqs.~(\ref{eq:dotpsi}) and $3H=-K$ we have
\begin{eqnarray}
0=-3\frac{\dot{a}}{a}{}_{(2)}\alpha+\frac{6\dot{\psi}}{\psi}(1-{}_{(2)}\alpha)
+O(\epsilon^4)\,.
\label{eq:Kdef2}
\end{eqnarray}
We see $\partial_t\psi=O(\epsilon^2)$ from this equation.

To summarize, the orders of magnitude of the basic metric quantities
are
\begin{eqnarray}
&&
\psi=O(1)\,,
\quad \beta^i=O(\epsilon)\,,
\quad \alpha-1=O(\epsilon^2)\,,
\quad \partial_t\tilde{\gamma}_{ij}=O(\epsilon^2)\,,
\quad \partial_t{\psi}=O(\epsilon^2)\,,
\quad \tilde{A}_{ij}=O(\epsilon^2)\,.
\end{eqnarray}
Actually, as for $\beta^i$, 
when we solve the equations to $O(\epsilon^2)$
in Appendix~\ref{sec:derivation}, we assume $\beta^i=O(\epsilon^3)$.
But since the choice of $\beta^i$ does not affect the 
temporal behavior, it does not affect the generality of the
solution. 

\section{Derivation of the general solution}
\label{sec:derivation}
%%%%%%%%%%%%%%%%%%%%%%%%%%%%%%%%%%%%%%%%%%%%%
%%subsec. Leading order equations
%%%%%%%%%%%%%%%%%%%%%%%%%%%%%%%%%%%%%%%%%%%%%
\subsection{The leading order solution}
We first derive the leading order solution for our basic variables.
Here, the "leading order" means not the lowest order in 
gradient expansion, but
the lowest order of each physical quantity. 
For example, from Eqs.~(\ref{eq:ordercount}),
the leading order of $\psi$ is $O(\epsilon^0)$, but the
 leading one of $\tilde A_{ij}$ is $O(\epsilon^2)$.

The $O(\epsilon^0)$ part of Eqs.~(\ref{eq:hamconst}) 
and (\ref{eq:dotK}) are
\begin{eqnarray}
&&\frac{1}{3}K^2(t)=3H^2=8\pi G\Bigl(\frac{\dot\phi^2}
{2}+V(\phi)\Bigr),
\label{eq:Fried1} \\ 
&&\dot K=-3\dot H=3H^2+4\pi G\left(\dot\phi^2-2V(\phi)\right).
\label{eq:Fried2}
\end{eqnarray}
These equations are indeed the Friedmann equations, and the leading solution
$\phi$ is that of a FLRW spacetime.

Setting $\beta^i=O(\epsilon^3)$, 
we substitute the order of magnitude evaluation of the variables
shown in Eq.~(\ref{eq:ordercount}) into the Klein-Gordon equation
and find
%%%%%%%%%%%%%%%%%%%%%%%%%%%%%%%%
%%% approximated KG eq.
%%%%%%%%%%%%%%%%%%%%%%%%%%%%%%%
\begin{eqnarray}
&&\ddot \phi+3H\dot\phi+\frac{dV}{d\phi}=0\,,
\label{eq:KG0} \\
&&{}_{(2)}\ddot \phi+3H{}_{(2)}\dot\phi
+\frac{d^2V}{d^2\phi}{}_{(2)}\phi-\dot\phi
\left({}_{(2)}\dot\alpha+3 H{}_{(2)}\alpha\right)-2\ddot\phi{}_{(2)}\alpha=0\,.
\label{eq:KG2}
\end{eqnarray}
 
%%%%%%%%%%%%%%%%%%%%%%%%%%%%%%%%%%%%
%%% approximated Einstein eq.
%%%%%%%%%%%%%%%%%%%%%%%%%%%%%%%%%%%%
The Hamiltonian and momentum constraint equations give
\begin{eqnarray}
&&\tilde{\gamma}^{ij}\tilde{D}_i\tilde{D}_j\psi=\frac{1}{8}
\tilde{\gamma}^{kl}\tilde{R}_{kl}\psi
-2\pi G\psi^5a^2
\left[-\dot\phi^2\postt\alpha+\dot\phi{}_{(2)}\dot\phi
+\frac{dV}{d\phi}{}_{(2)}\phi\right]+O(\epsilon^4)\,,
\label{eq:aphamconst} \\
&&\tilde{D}^j(\psi^6\tilde{A}_{ij})
=-8\pi G\psi^6\dot\phi \,\partial_j({}_{(2)}\phi)+O(\epsilon^5)\,,
\label{eq:apmomconst}
\end{eqnarray}
where $\tilde{R}_{ij}$ is the Ricci tensor with respect 
to $\tilde{\gamma}_{ij}$, and $\tilde{D}_i$ is the covariant derivative 
with respect to $\tilde{\gamma}_{ij}$.

The evolution equations for the spatial metric give
\begin{eqnarray}
&&\frac{6\partial_t{\psi}}{\psi}-3H(\alpha-1)=D_k \beta^k,
\label{eq:psidot} \\
&&(\partial_t-\beta^k\partial_k)\tilde{\gamma}_{ij}=-2\tilde{A}_{ij}
+\tilde{\gamma}_{ik}\partial_j\beta^k+\tilde{\gamma}_{jk}\partial_i\beta^k
-\frac{2}{3}\tilde{\gamma}_{ij}\partial_k\beta^k+O(\epsilon^4)\,,
\label{eq:gammadot}
\end{eqnarray}
while the evolution equations for the extrinsic curvature give
\begin{eqnarray}
&&\partial_t \tilde{A}_{ij}+3H\tilde{A}_{ij}
=\frac{1}{a^2\psi^4}[R_{ij}-\frac{\gamma_{ij}}{3}R]+O(\epsilon^4)\,,
\label{eq:Adot} \\
&&0=3H^2{}_{(2)}\alpha
+8\pi G\left[-\left(\dot\phi^2+V(\phi)\right){}_{(2)}\alpha
+2\dot\phi{}_{(2)}\dot\phi-\frac{dV}{d\phi}{}_{(2)}\phi
\right]\,.
\label{eq:alpha2}
\end{eqnarray}

{}First, for $\psi$, we have
\begin{eqnarray}
\psi=\prez L(x^i)+O(\epsilon^2),
\label{eq:psizero}
\end{eqnarray}
where $\prez L$ is an arbitrary function of the spatial coordinates.
And from Eqs.~(\ref{eq:Adot}) and (\ref{eq:psizero}) 
together with the assumption $\partial_t{\tilde\gamma}_{ij}=O(\epsilon^2)$,
we obtain
\begin{eqnarray}
&&\tilde\gamma_{ij}=\prez f_{ij}(x^k)+O(\epsilon^2)\,,
\label{eq:prezfij}\\
&&\tilde{A}_{ij}=\frac{\pret F_{ij}}{a^3}\int_{t_*}^t
 a(t^{\prime})dt^{\prime}+\frac{\pret C_{ij}(x^k)}{a^3}+O(\epsilon^4),
\label{eq:Asol}\\
&&\quad \pret F_{ij}\equiv \frac{1}{\prez L^4}\left[\pret \bar{R}_{ij}
+\pret R^L_{ij}-\frac{\prez f_{ij}}{3}\,
\prez f^{kl}\left(\pret \bar{R}_{kl}+\pret R^L_{kl}\right)\right]\,,
\label{eq:Fijdef}\\
&&\quad \pret R^L_{ij}\equiv -\frac{2}{\prez L}\bar{D}_i\bar{D}_j\prez L
-\frac{2}{\prez L}\prez f_{ij}\bar{\Delta}
\prez L+\frac{6}{\prez L^2}\bar{D}_i\prez L\bar{D}_j\prez L
-\frac{2}{\prez L^2}\prez f_{ij}\bar{D}_k\prez L\bar{D}^k\prez L\,,
\label{eq:RLdef}
\end{eqnarray}
where $t_*$ is an arbitrary reference time,
$\prez f_{ij}$ is an arbitrary and symmetric tensor of the spatial coordinates,
$\bar{R}_{ij}$ is the Ricci tensor with respect to $\prez f_{ij}$,
$\bar D_i$ is the covariant derivative with respect to $\prez f_{ij}$,
$\bar \Delta$ is the Laplacian with respect to $\prez f_{ij}$, 
and $\pret C_{ij}$ is an arbitrary, symmetric and traceless tensor 
which depends only on the spatial coordinates.

%%%%%%%%%%%%%%%%%%%%%%%%%%%%%%%%%
%% Sec. Subsection 
%%  solution of curvature perturbation in O(\epsilon^2)
%%%%%%%%%%%%%%%%%%%%%%%%%%%%%%%%
\subsection{The solution to $O(\epsilon^2)$ in gradient expansion}
Now we consider the general solution valid to $O(\epsilon^2)$
in gradient expansion. As we have seen in the previous subsection,
among the basic variables we have introduced,
the only quantities whose leading order terms are
lower than $O(\epsilon^2)$ are $\psi$, $\tilde\gamma_{ij}$, $\alpha$, and $\phi$.
Hence what we have to do is to evaluate the next order terms of these variables.
%%%%%%%%%%%%%%%%%%%%%%%%%%%%%%%%%%%%%%
%% \phi solution in O(\epsilon^2)
%%%%%%%%%%%%%%%%%%%%%%%%%%%%%%%%%%%%%%

First let us consider $\tilde\gamma_{ij}$.
Substituting~Eq.~(\ref{eq:Asol}) in Eq.~(\ref{eq:gammadot}), 
and choosing $\beta^i=O(\epsilon^3)$,
we obtain the general solution for $\tilde \gamma_{ij}$,
\begin{equation}
\tilde\gamma_{ij}=\prez f_{ij}(x^k)
-2\pret F_{ij}(x^k)\int^t \frac{dt'}{a^3(t')}\int^{t'} a(t'')dt''
-2\pret C_{ij}(x^k)\int^t\frac{dt'}{a^3(t')}+O(\epsilon^4)
+\pret f_{ij}(x^k)\,,
\end{equation}
where $\pret f_{ij}$ is an arbitrary and symmetric tensor
which depends only on the spatial coordinates. We may absorb it
into $\prez f_{ij}$ without loss of generality. Thus we have
\begin{equation}
\tilde\gamma_{ij}=\prez f_{ij}(x^k)
-2\pret F_{ij}(x^k)\int^t \frac{dt'}{a^3(t')}\int^{t'} a(t'')dt''
-2\pret C_{ij}(x^k)\int^t\frac{dt'}{a^3(t')}+O(\epsilon^4)\,,
\label{eq:tildegammasol}
\end{equation}
Here, we note that $\prez f_{ij}$ is not completely arbitrary,
but its determinant must be unity, $\det(\prez f_{ij})=1+O(\epsilon^4)$.

To obtain the other variables, we first consider the solution
for $\pret\phi$. We express $\pret\alpha$ in terms of $\pret\phi$
by using Eq.~(\ref{eq:alpha2}). We obtain
\begin{eqnarray}
\pret\alpha=\frac{2}{\dot\phi^2}
\left(2\dot\phi\,\pret\dot\phi-\frac{dV}{d\phi}\,\pret\dot\phi\right)
\label{eq:alpha2exp}
\end{eqnarray}
Inserting this into the field equation (\ref{eq:KG2}), 
we obtain a closed equation for $\pret\phi$,
\begin{eqnarray}
\pret\ddot\phi-\frac{H\dot\phi+2\,dV/d\phi}{\dot\phi}\pret\dot\phi
-\frac{\dot\phi\,d^2V/d\phi^2+2HdV/d\phi}{\dot\phi}\pret\phi=0\,.
\label{eq:phi2eq}
\end{eqnarray}
Although this equation looks difficult to solve at first glance,
it turns out that it can be analytically solved.
Here, we note that because the lowest order scalar field is 
only a function of time, the above equation is exactly the same as 
the one in the long wavelength limit of linear theory.
Then, it is known that there exits a particular solution that satisfies
\begin{equation}
\dot\phi\,\pret\dot\phi_d-\frac{dV}{d\phi}\pret\phi_d=0\,,
\label{eq:PhiDecaymode}
\end{equation}
where the subscript $d$ is attached since it corresponds to a decaying
mode solution in the linear limit.
This may be integrated easily to give
\begin{eqnarray}
\pret\phi_d\propto v(t)\equiv
\exp\left[\int^tdt\frac{dV/d\phi}{\dot\phi}\right]
=\frac{\dot\phi(t_*)}{a^3(t)\dot\phi(t)}\,,
\label{eq:vsoldef}
\end{eqnarray}
where we have normalized the amplitude so that $v=1/a^3(t_*)$ at $t=t_*$.

Once we know a particular solution, the other independent solution, $u(t)$,
can be found by the use of the Wronskian. For two independent solutions
$u$ and $v$ of Eq.~(\ref{eq:phi2eq}), the Wronskian,
\begin{eqnarray}
W=\dot u\,v-\dot v\,u\,,
\end{eqnarray}
satisfies
\begin{eqnarray}
\dot W+b(t)W=0\,,
\end{eqnarray}
where $b(t)$ is the coefficient of $\pret\dot\phi$ in Eq.~(\ref{eq:phi2eq}),
\begin{eqnarray}
b\equiv-\frac{H\dot\phi+2\,dV/d\phi}{\dot\phi}\,.
\end{eqnarray}
Thus we obtain
\begin{eqnarray}
\dot u\,v-\dot v\,u
=W\propto \exp\left[-\int^tb(t)dt\right]=\frac{a}{(a^3\dot\phi)^2}
\propto av^2\,.
\end{eqnarray}
Then since
\begin{eqnarray}
\frac{d}{dt}\left(\frac{u}{v}\right)=\frac{\dot u\,v-\dot v\,u}{v^2}=\frac{W}{v^2}\,,
\end{eqnarray}
we readily find
\begin{eqnarray}
u=v\int^t_{t_*}\frac{W}{v^2}dt'
=\frac{\dot\phi(t_*)}{a^3(t)\dot\phi(t)}\int_{t_*}^ta(t')dt'\,,
\label{eq:usoldef}
\end{eqnarray}
where we have normalized $u$ so that $(u/v)\dot{}=a(t_*)$ at $t=t_*$.
Thus, the general solution for $\pret\phi$ is given by
\begin{eqnarray}
\pret\phi=C(x^k)u(t)+D(x^k)v(t)\,.
\label{eq:phisol}
\end{eqnarray}

Given the general solution for $\pret\phi$, the remaining variables 
$\pret\alpha$ and $\pret\psi$ are
determined as follows. Equation~(\ref{eq:alpha2exp}) gives
the solution for $\pret\alpha$ as
\begin{eqnarray}
\pret\alpha
&=&\frac{2}{\dot\phi^2}\left[\left(2\dot\phi\,\dot u-\frac{dV}{d\phi}\,u\right)C(x^k)
+\left(2\dot\phi\,\dot v-\frac{dV}{d\phi}\,v\right)D(x^k)\right]
\nonumber\\
&=&\frac{2\dot\phi_*}{\dot\phi^3a^3}
\left[\left(2a\,\dot\phi+\frac{dV}{d\phi}\int_{t_*}^ta(t')dt'\right)C(x^k)
+\frac{dV}{d\phi}D(x^k)\right]\,.
\label{eq:alphasol}
\end{eqnarray}
In terms of this solution, Eq.~(\ref{eq:psidot}) gives 
the $O(\epsilon^2)$ part of $\psi$ as
\begin{eqnarray}
\dot\psi&=&\frac{H}{2}\prez L\,\pret\alpha\,,
\end{eqnarray}
where we have set $\beta^i=O(\epsilon^3)$.
Integrating this equation, we obtain
\begin{equation}
\psi=\prez L\left[1+\frac{1}{2}\int_{t_*}^t\pret\alpha \,Hdt'\right]+\pret L(x^k)\,,
\label{eq:psisol}
\end{equation}
where $\pret L(x^k)$ is an arbitrary spatial function 
of $O(\epsilon^2)$, which we may absorb into $\prez L$ without loss of generality.

%%%%%%%%%%%%%%%%%%%%%%%%%%%%%%%%%%%%%%%%%
%% Constraint equations 
%%%%%%%%%%%%%%%%%%%%%%%%%%%%%%%%%%%%%%%%%
Up to now we have not considered the Hamiltonian and momentum
constraint equations. The constraint equations will relate the
quantities $\prez L$, $\prez f_{ij}$, $\pret C_{ij}$, $\pret C$ and $\pret D$.
First let us focus on the Hamiltonian constraint, Eq.~(\ref{eq:aphamconst}).
Using Eqs.~(\ref{eq:alpha2}) and (\ref{eq:psizero}), it becomes
\begin{eqnarray}
\frac{1}{a^2\prez L^5}
\left[-8\prez f^{ij}\bar{D}_i\bar{D}_j\prez L
+\prez f^{kl}\pret\bar{R}_{kl}\prez L\right]=
48\pi G\left[-\dot\phi{}_{(2)}\dot\phi+\frac{dV}{d\phi}{}_{(2)}\phi\right].
\label{eq:Hamconst3} 
\end{eqnarray}
Comparing this equation with Eq.~(\ref{eq:PhiDecaymode}),
we see that the right-hand side vanishes for the decaying mode
solution $v$. Thus inserting the general solution~(\ref{eq:phisol}) into the above,
we obtain
\begin{eqnarray}
\frac{1}{a^2\prez L^5}
\left[-8\prez f^{ij}\bar{D}_i\bar{D}_j\prez L
+\prez f^{kl}\pret\bar{R}_{kl}\prez L\right]
=-48\pi G\frac{\dot\phi(t_*)}{a^2}C(x^k)\,.
\end{eqnarray}
Thus $D(x^k)$ is not constrained by the Hamiltonian, while $C(x^k)$ is
determined in terms of $\prez L$ and $\prez f_{ij}$ as
\begin{eqnarray}
\pret C&=&\frac{-1}{48\pi G\prez L^5\dot\phi_{\ast}}
\bigl[-8\prez f^{ij}\bar{D}_i\bar{D}_j\prez L
+\prez f^{kl}\pret\bar{R}_{kl}\prez L
\bigr]+O(\epsilon^4)
\nonumber\\
&=&\frac{-1}{48\pi G\prez L^4 \dot\phi_{\ast}}
\prez f^{kl}\left[\pret R^L_{kl}+\pret \bar{R}_{kl}\right]+O(\epsilon^4)
\nonumber\\
&=&\frac{-1}{48\pi G\dot\phi_{\ast}}
\,R\left[L^4f\right]+O(\epsilon^4)\,,
\label{eq:Cdef}
\end{eqnarray}
where $\dot\phi_*=\dot\phi(t_*)$ and $R\left[L^4f\right]$ is 
the Ricci scalar of the metric $\prez L^4\prez f_{ij}$.

The $O(\epsilon^3)$ part of the momentum constraint~(\ref{eq:apmomconst}) yields
\begin{eqnarray}
\bar{D}^j\left[\prez L^6\pret F_{ij}\right]
&=&-8\pi G\prez L^6\dot\phi_{\ast}\partial^j\left(\pret C(x^k)\right)\,,
 \label{eq:DF} \\
\bar{D}^j\left[\prez L^6\pret C_{ij}\right]
&=&-8\pi G\prez L^6\dot\phi_{\ast}\partial^j\left(\pret D(x^k)\right)\,.
\label{eq:CC} 
\end{eqnarray}
The latter equation implies $\pret D$ is not arbitrary
but expressed in terms of $\prez L$, $\prez f_{ij}$ and $\pret C_{ij}$,
while the former equation is found to be consistent with
Eq.~(\ref{eq:Cdef}). This consistency is a result of the 
Bianchi identities.

%%%%%%%%%%%%%%%%%%%%%%%%%%%%%%%%%
%% thebibliography environment 
%%%%%%%%%%%%%%%%%%%%%%%%%%%%%%%%%

\end{document}